\documentstyle[12pt]{article}
\begin{document}

\baselineskip=0.50truecm

\title{Linear Potentials and Galactic Rotation Curves - Detailed Fitting
\footnote
{UCONN 96-02, February, 1996}}

\author{\normalsize{Philip D. Mannheim} \\
\normalsize{Department of Physics,
University of Connecticut, Storrs, CT 06269} \\ 
\normalsize{mannheim@uconnvm.uconn.edu} \\
\normalsize{and} \\
\normalsize{Jan Kmetko} \\
\normalsize{Department of Physics,
Berea College, Berea, KY 40404} \\}

\date{}

\maketitle

\begin{abstract}
We continue our study of the astrophysical implications of the linear 
potential $V(r)=-\beta c^2/r+\gamma c^2 r/2$ associated with fundamental 
gravitational sources in the conformal invariant fourth order theory of 
gravity which has recently been advanced by Mannheim and Kazanas as a 
candidate alternative to the standard second order Einstein theory. We 
provide fitting to the rotation curves of an extensive and diverse set of 
11 spiral galaxies whose data are regarded as being particularly 
reliable. Without the assumption of the existence of any dark matter 
the model is found to fit the shapes of the rotation curves extremely well, 
but with a pattern of normalizations which proves to be very instructive. 
\end{abstract}

\section{Introduction}

During the last few years Mannheim and Kazanas (Mannheim 1990, 1992, 1993a, b,
1994, 1995a, b, c; Mannheim and Kazanas 1989, 1991, 1994; Kazanas and Mannheim
1991) have been exploring conformal gravity (viz. gravity based on invariance
of the geometry under any and all local conformal stretchings of the form 
$g_{\mu \nu} (x) \rightarrow \Omega (x) g_{\mu \nu} (x)$) as a 
covariant candidate alternative to the standard Newton-Einstein gravitational
theory. Their study has entailed both the examining of the formal structure of
the theory and the identification of its possible observational astrophysical 
implications. In particular they found (Mannheim and Kazanas 1989; see also 
Riegert 1984) the most general, exact all order metric exterior to a static, 
spherically symmetric source such as a star in the theory, viz. (in standard 
static coordinates)
\begin{equation}
-g_{00}= 1/g_{rr}=1-\beta(2-3 \beta \gamma )/r - 3 \beta \gamma               
+ \gamma r - kr^2                                                     
\label{Eq. (1.1)}
\end{equation}
where $\beta, \gamma$ and $k$ are three integration constants. Subsequently,
they also found (Mannheim and Kazanas 1994) the associated exact interior 
solution and established its consistency with the exterior one, while also 
showing that only the coefficients of the $r$ and $1/r$ terms in the general  
Eq. (\ref{Eq. (1.1)}) were actually related to properties of the source. As we 
can thus see, the metric of Eq. (\ref{Eq. (1.1)}) not only generalizes the 
potential of Newtonian gravity, it also generalizes the Schwarzschild solution 
of Einstein gravity as well, so that (for an appropriate choice of the 
coefficients of the $r$ and $1/r$ terms) the conformal theory can nicely recover 
the Newtonian potential and its familiar Einstein relativistic corrections on the 
small distance scale associated with the solar system, and then depart from the 
standard theory on the much larger distance scale associated with galaxies, this 
being precisely the distance scale where the standard Newton-Einstein theory 
can apparently only survive if galaxies contain copious amounts of dark matter. 
And, moreover, the first fitting (Mannheim 1993a) of the linear potential of  
Eq. (\ref{Eq. (1.1)}) to a small set of four characteristic galaxies showed 
that the conformal theory could actually account for the relevant rotation 
curve data without the need to invoke dark matter at all.    
  
As regards the actual possible existence of galactic dark matter, we note that 
neither the vigorous dark or faint matter searches of the OGLE (Udalski et al 
1993, 1994), MACHO (Alcock et al 1993) and EROS (Aubourg et al 1993) 
gravitational microlensing collaborations nor those using the unprecedented 
optical sensitivity now available to the recently refurbished Hubble Space 
Telescope (Bahcall et al 1994) have so far been able to confirm the 
presence of the huge spherical dark matter galactic halo required of the 
standard theory.  At the very minimum one can say that these searches have 
certainly not yet achieved their intended goal of validating the standard 
picture, while at the maximum one can say that they have even actually thrown 
the entire picture into doubt. While it is of course far too early to 
contemplate abandoning the standard paradigm, nonetheless the current 
observational situation does demand a critical reappraisal of its two key 
components, namely the presumed existence of dark matter and the assumed 
validity of the Newton-Einstein gravitational theory on distance scales much 
larger than the solar system one on which it was first established.
Moreover, the very assumption of the continuing validity of the standard 
theory on these much larger galactic distance scales represents a so far 
unjustified and possibly even dangerous extrapolation; with the very need for 
dark matter possibly even being an indicator that such an extrapolation is not 
in fact reliable. Since conformal gravity also reproduces the standard solar 
system wisdom while leading to a very different galactic extrapolation, it 
would thus appear to be a legitimately motivated gravitational theory whose 
eventual ultimate status can only be ascertained through consideration of its 
observational consequences. The present paper therefore sets out to explore 
further the observational implications of conformal gravity by applying it to 
a quite extensive and diverse 11 galaxy rotation curve sample. While this 
would appear to be a straightforward enough procedure, as we shall actually 
see, the fitting we present in this paper will lead us to a somewhat 
unanticipated conclusion.   
 
\section{The Model and the Data Sample}

Given the metric of Eq. (\ref{Eq. (1.1)}), we can take the individual stellar
potentials of each of the $N^{*}$ stars in a galaxy to be of the form
\begin{equation}
V^{*}(r)=-\beta^{*}c^2/r+ \gamma^{*}c^2r/2                                                  
\label{Eq. (2.1)}
\end{equation}
to give a potential which is then to be integrated over the galactic matter
distribution in the standard non-relativistic Newtonian way. For the matter 
distribution we take as its components the visible stars and the detected HI 
gas. Since optically only the luminosity surface density is detected, for the 
stars we shall assume that their matter surface density distribution is given in 
shape by the detected surface brightness but normalized to it with a (position 
independent) mass to light ratio $M/L$, a ratio which is however allowed to vary 
from one galaxy to the next. For the HI gas the absolute mass normalization is 
inferrable from the data once the distance to any galaxy is determined. For many
spiral galaxies, after the extracting out of any possible central spheroidal
bulge, the remaining optical disk surface brightness can be well approximated by 
the separable product $I(R)f(z)$ where $R$ is the distance from the galactic 
center within the galactic plane and $z$ is the height above the galactic 
plane; with the fall off of the $I(R)$ intensity in the plane typically being 
an exponential with a scale length $R_0=1/\alpha$, and with the fall off 
perpendicular to the plane typically giving the disk a thickness of a form such 
as the $f(z)=sech^2(z/z_0)/2z_0$ profile originally found by van der Kruit and 
Searle (1981) in studies of edge on galaxies. (While many fits are made in the 
literature using an infinitesimally thin disk, for completeness we shall include 
this thickness factor here, but since we shall assume the typical 
$z_0/R_0=0.2$ ratio for all the optical disks in our 11 galaxy sample, the 
effect of the thickness turns out to only be numerically significant in the 
inner region, and is thus of no consequence for the outer galactic region where 
the luminous Newtonian prediction faces all its current difficulties.)  With the 
use of the Bessel functions which are characteristic of axial symmetry, it is 
possible (Mannheim 1995a) to obtain closed form expressions for the rotational 
velocities of orbits in the galactic plane. Thus for an infinitesimally thin 
disk with surface matter distribution $\Sigma(R)=\Sigma_0$exp$(-R/R_0)$ and a 
total of $N^{*}=2 \pi \Sigma_0 R_0^2$ stars each with potential $V^{*}(r)$, we 
obtain (Mannheim 1995a) for the complete galactic potential of the disk 
\begin{eqnarray}
rV^{\prime}(r)=
(N^{*}\beta^{*} c^2\alpha^3 r^2/2)[I_0(\alpha r/2)K_0(\alpha r/2)-
I_1(\alpha r/2)K_1(\alpha r/2)]
\nonumber \\
+(N^{*}\gamma^{*} c^2r^2\alpha/2)I_1(\alpha r/2)K_1(\alpha r/2)                                                  
\label{Eq. (2.2)}
\end{eqnarray}
\noindent
Similarly, for a disk with an additional $sech^2(z/z_0)/2z_0$ thickness we 
obtain (Mannheim 1995a)
\begin{eqnarray}
rV^{\prime}(r)=
(N^{*}\beta^{*} c^2\alpha^3 r^2/2)[I_0(\alpha r/2)K_0(\alpha r/2)-
I_1(\alpha r/2)K_1(\alpha r/2)]
\nonumber \\
-N^{*} \beta^{*} c^2\alpha^3 r\int_0^\infty dk {k^2J_1(kr)z_0\beta(1+kz_0/2) 
\over (\alpha^2+k^2)^{3/2}} 
\nonumber \\
+N^{*}\gamma^{*} c^2\alpha^3 r \int_{0}^{\infty}dk
(1-kz_0\beta(1+kz_0/2))
\nonumber \\
\times~ \left( -{2rJ_0(kr) \over (\alpha^2+k^2)^{3/2}}
+{3\alpha^2rJ_0(kr) \over (\alpha^2+k^2)^{5/2}}
-{r^2kJ_1(kr) \over 2(\alpha^2+k^2)^{3/2}}
+{9kJ_1(kr) \over 2(\alpha^2+k^2)^{5/2}}
-{15\alpha^2kJ_1(kr) \over 2(\alpha^2+k^2)^{7/2}} \right)
\nonumber \\
+N^{*}\gamma^{*} c^2\alpha^3 r \int_{0}^{\infty}dk{kJ_1(kr)
\over 2(\alpha^2+k^2)^{3/2}} 
{d^2 \over dk^2}\left( kz_0\beta(1+{kz_0 \over 2}) 
\right)                                                  
\label{Eq. (2.3)}
\end{eqnarray}
where $\beta(x)$ is the polygamma function  
\begin{equation}
\beta(x)=\int_0^1{t^{x-1} \over (1+t)} 
\label{Eq. (2.4)}
\end{equation}
\noindent
Since the $\beta(x)$ function and its derivatives converge very rapidly to 
their asymptotic values as their arguments increase, the $k$ integrations in 
Eq. (\ref{Eq. (2.3)}) converge very rapidly thus making numerical evaluation 
very simple. For galaxies which also possess a spherical bulge, its contribution 
to the total galactic potential can be expressed as a closed form function of 
the projected surface matter distribution $I(R)$ to yield (Mannheim 1995a)
\begin{eqnarray}
rV^{\prime}(r)=
{4 \beta^{*} c^2\over r}\int_r^{\infty}dRRI(R)
\left[ arcsin \left({r \over R} \right) -{r \over (R^2-r^2)^{1/2}}\right] 
\nonumber \\
+{2 \pi \beta^{*} c^2\over r} \int_0^rdRRI(R) +
{\gamma^{*} c^2 \pi \over 2r} \int_0^rdRRI(R)(2r^2-R^2)
\nonumber \\
+{\gamma^{*} c^2  \over r}\int_r^{\infty}dRRI(R)
\left[ (2r^2-R^2)arcsin \left({r \over R} \right) +r(R^2-r^2)^{1/2}\right] 
\label{Eq. (2.5)}
\end{eqnarray}
\noindent
Equations (\ref{Eq. (2.2)}), (\ref{Eq. (2.3)}) and (\ref{Eq. (2.5)})
thus define the model.

For the data sample we follow Begeman, Broeils and Sanders (1991) and use their
chosen 11 galaxy sample, a sample which satisfies the no less than 8 independent 
and demanding selection criteria which they established. Since the resulting 
sample of galaxies is then found to contain galaxies which range in luminosity 
by a factor of more than 1000, the sample should indeed be regarded as 
representative of rotation curve systematics. The actual data for this sample 
consist of optical disk photometry, HI gas rotational velocities, and HI gas 
surface density profiles $\sigma_{HI}(R)$. In Table I we list the complete 
sample, the adopted distances to the galaxies (normalized to a Hubble distance 
$H_0=75$ km s$^{-1}$ Mpc$^{-1}$), the associated luminosities, optical disk 
scale lengths $R_0$,\footnote{The two largest galaxies in our sample are found 
to also contain appreciable spheroidal bulges. For NGC 7331 the bulge to disk 
luminosity is 1.4 with the bulge truncating at $1.42^{\prime}$; while for 
NGC 2841 the bulge to disk luminosity is 0.4 with the bulge truncating at 
$50^{\prime\prime}$.} total HI gas masses,\footnote{To account for some missing 
21 cm line flux at the VLA where the NGC 3109 observations of Jobin and 
Carignan (1990) were made, for fitting purposes we shall follow Begeman, Broeils 
and Sanders (1991) and multiply the total NGC 3109 HI mass reported in Table I by 
an additional factor of 1.67} and indicate the data sources.\footnote {In Table 
I the data sources are denoted by: 1, Carignan and Freeman (1988), Carignan and 
Beaulieu (1989); 2, Lake, Schommer and van Gorkom (1990); 3, Broeils (1992); 4, 
Jobin and Carignan (1990); 5, Carignan, Sancisi and van Albada (1988); 6, 
Wevers, van der Kruit and Allen (1986); 7, Begeman (1987, 1989); 8, Kent 
(1987).} Table I is basically as given by Begeman, Broeils and Sanders in their 
paper, save that for NGC 2841 we have used the adopted distance favored 
subsequently by Sanders and Begeman (1994) in a follow up paper. In order to be 
able to use the simple formula of Eq. (\ref {Eq. (2.2)}) for the HI gas 
contribution (which we take to have no thickness), we have decomposed the gas 
surface densities into sums of exponentials\footnote{The DDO 154 
gas profile is fit by $\sigma_{HI}(R)=31.60$exp$(-R/1.42)-25.70$exp$(-R/1.08)$ 
in units of $M_{\odot}$ pc$^{-2}$ with $R$ being in arcminutes. DDO 170
is fit by $\sigma_{HI}(R)=22.94$exp$(-R/0.73)-18.08$exp$(-R/0.55)$. NGC 1560
is fit by $\sigma_{HI}(R)=86.33$exp$(-R/2.64)-74.75$exp$(-R/2.28)$. NGC 3109
is fit by $\sigma_{HI}(R)= 6.16$exp$(-R/7.90)- 2.86$exp$(-R/1.05)$. UGC 2259
is fit by $\sigma_{HI}(R)=37.53$exp$(-R/0.74)-48.11$exp$(-R/0.49)+
14.66$exp$(-R/0.26)$. NGC 6503 is fit by 
$\sigma_{HI}(R)=64.76$exp$(-R/2.96)-58.90$exp$(-R/2.80)$. NGC 2403 is fit by 
$\sigma_{HI}(R)=58.47$exp$(-R/4.16)-76.02$exp$(-R/2.47)+
25.88$exp$(-R/1.39)$. NGC 3198 is fit by 
$\sigma_{HI}(R)=36.99$exp$(-R/2.23)+34.58$exp$(-R/0.87)-
68.18$exp$(-R/1.21)$. NGC 2903 is fit by
$\sigma_{HI}(R)= 10.89$exp$(-R/3.27)- 6.60$exp$(-R/0.91)$. NGC 7331 is fit by
$\sigma_{HI}(R)= 48.10$exp$(-R/1.89)- 47.50$exp$(-R/1.22)$, and NGC 2841 is	
fit by $\sigma_{HI}(R)= 6.04$exp$(-R/4.10)- 5.94$exp$(-R/0.89)$. As regards
these fits we note in passing that the dominant exponential in each fit 
typically has a scale length a factor 3 or so times larger than that of the 
associated optical galactic component.} (and additionally for fitting purposes 
we multiply the HI gas profile by a factor 1.4 to account for primordial 
helium). Beyond the issue of simplicity of calculation provided by such an
exponential decomposition, we note that unlike the Newtonian $1/r$ potential, 
the linear potential has the property that the total galactic potential at any 
given point is sensitive to the presence of any matter exterior to it. The 
fitting of the gas profile (and also the optical profile) by exponentials then 
provides for well defined continuations of these densities beyond the detected 
region. With the decomposition of the gas profile the input parameters to the 
model are now completely specified.

\section{The Fitting}

While the Newtonian sector of the model is well understood with just one free 
parameter per galaxy, viz. the disk mass to light ratio $(M/L)_D$, (with a 
possible independent bulge mass to light ratio as well when relevant), the 
linear sector of the theory is not yet as well understood, with the stars and 
gas possibly having differently parameterized linear terms. Consequently, in 
order to restrict the number of free parameters in the linear sector to be the 
same number as that in the Newtonian one, we shall impose relations between the 
$\gamma^{*}$ and $\gamma_{gas}$ parameters needed for Eqs. (\ref{Eq. (2.2)}) and 
(\ref{Eq. (2.3)}). Then by treating $\gamma^{*}$ as a free parameter which we 
(initially at least) allow to vary from one galaxy to the next, the model is
thus set up with two free parameters per galaxy. Since the standard dark matter
spherical halo model comes with three free parameters per galaxy, our model is 
already more tightly constrained than the standard one. As to possible relations 
between $\gamma_{gas}$ and $\gamma^{*}$, two candidates ones were originally 
suggested in the first fitting of the linear potential model (Mannheim 1993a), 
namely $\gamma_{gas}=\gamma^{*}$ and $\gamma_{gas}=0$, and so we follow them 
here. The fitting to the full 11 galaxy set when the condition 
$\gamma_{gas}=\gamma^{*}$ is imposed on each galaxy is presented in Figure (1) 
with the associated output parameters being given in Table (2);\footnote{ For
NGC 7331 and NGC 2841 the fitted bulge mass to light ratios are  given by 
1.03 and 0.72 respectively.} with the fitting associated with $\gamma_{gas}=0$ 
being presented in Figure (2) and Table (3).\footnote{ For NGC 7331 and NGC 2841 
the fitted bulge mass to light ratios are  given by 1.40 and 0.22 respectively.} 
In Tables (2) and (3) we have listed not only the fitted values of $(M/L)_D$ and 
$\gamma^{*}$, but also we have given the total linear potential coefficient 
$\gamma_{gal}$ summed over the entire galaxy (viz. the total 
$N^{*}_D\gamma^{*}+N^{*}_B\gamma^{*}+N_{gas}\gamma_{gas}$ for disk, bulge and 
gas combined), the gamma to light ratio $(N^{*}\gamma^{*}/L)_D$ for the disk,
and the magnitude of the dimensionless ratio $\gamma^{*} R^2_0/\beta^{*}$ 
which controls the relative strengths of the disk Newtonian and linear terms in 
Eq. (\ref{Eq. (2.2)}). By determining the fitted values of all of these 
quantities, we generate a base big enough to enable us to search for any 
possible regularities in the fits. 

As we can see from Figures (1) and (2), our model is able to fit the shapes of 
the rotation curves remarkably well, something which is quite non-trivial in 
and of itself, and especially so for a theory with asymptotically rising 
rather than asymptotically flat rotational velocities; with the ultimate 
required rise in the rotation curves being nicely postponable in the high 
luminosity spirals beyond the flat rotation curve regions they currently 
exhibit. Additionally, since the quality of the fits in the two figures is
comparable, we see that there is a fairly broad region in parameter space which 
gives acceptable fitting. (In fact variable $\gamma^{*}$ fits of a comparable
quality can even be obtained when $\gamma_{gas}$ is allowed to take a common 
non-zero value for all of the 11 galaxies in the sample, viz. any value in the
range from $\gamma_{gas}=0$ to $\gamma_{gas}=1.5 \times 10^{-40}$ cm$^{-1}$ per 
unit solar mass of gas.) As a check on our numerical work, we note that 
equivalent fitting to the same galaxy sample with a comparable set of output 
parameters has also been obtained by Carlson and Lowenstein (1996) whose work 
was performed contemperaneously with our own, with each of the two studies thus 
confirming the other. 
 
Examination of the fits presented in Tables (2) and (3) reveals two immediately   
striking features. Firstly, we find that the fitted values of $\gamma^{*}$ for 
our 11 galaxies turn out to be nowhere near close to each other in magnitude 
(this being the a priori Newtonian expectation of course), with the fitted 
values in fact showing a quite marked decrease with increasing luminosity.
\footnote{The fitted $\gamma^{*}$ values obtained for the gas rich galaxy NGC 
3109 show a deviation from the general decreasing trend found for the other 10 
galaxies, a feature which is not necessarily of significance since, as we had 
noted earlier, the amount of gas used in the NGC 3109 fits was fixed using only 
an estimate of the actual amount of flux missing from the VLA data.} And, 
secondly, and quite unexpectedly, we also find that, rather than 
$\gamma^{*}$ being universal, instead it is the 
total $\gamma_{gal}$ summed over all the stars and the gas in each galaxy which 
turns out to be universal, universal in fact to within a factor of three 
according to Tables (2) and (3), and this despite a variation of more than 1000 
in luminosity throughout the 11 galaxy sample.\footnote{We note that this same 
universality is also manifest in the original linear potential fits of Mannheim 
(1993a), though with only four galaxies having been fitted there, it was  
difficult to assess the generality of the finding. Now with the large galaxy 
sample of the present work, the generality of this regularity becomes apparent, 
a regularity which is also manifest in the independent study made by Carlson and 
Lowenstein (1996).} \footnote{The specific mathematical reason why the fits 
actually lead to universal $\gamma_{gal}$ rather than universal $\gamma^{*}$ in 
the first place was identified by Mannheim (1995b) who pointed out that for each 
galaxy in the sample the centrifugal acceleration $v^2/R$ at the data point 
furthest from its center was numerically very close in magnitude to 
$\gamma_{gal}c^2/2$, to thereby yield a universality which the linear potential 
fits have no choice but to respect, with the universality we then find for 
$\gamma_{gal}$ thus not being merely an artifact of the fitting procedure.} 
Given this universality for $\gamma_{gal}$, it then follows that the disk gamma 
to light ratio $(N^{*}\gamma^{*}/L)_D$ must fall sharply with increasing 
luminosity, this of course being in marked contrast to the disk mass to light 
ratios for the same fits which show little variation with luminosity, just as 
they show little variation in the standard dark matter fits. Since the higher 
luminosity galaxies in our sample all have a (close to) common mean or central 
surface brightness (the Freeman limit value  $\Sigma_0^F$ first identified for 
regular spirals by Freeman 1970), the universality of $\gamma_{gal}$ also 
translates into the (near) universality of the dimensionless ratio 
$\gamma^{*} R^2_0/\beta^{*}$ exhibited in the fitting.

The universality that we find for the total $\gamma_{gal}$ is as puzzling as it 
is striking, and quite at variance with the naive Newtonian expectation of a 
total $\gamma_{gal}$ which should grow with luminosity just like the coefficient
of the total galactic Newtonian term. In order to both define and sharpen the 
exact nature of the puzzle, we recall that in our fitting we determined the 
total galactic potential simply by naively integrating the stellar potential 
$V^{*}(r)$ over the detected luminosity distribution as normalized with 
a position independent mass to light ratio. However, the assumption of a 
position independent mass to light ratio is certainly invalid, at least in 
principle, since the luminosities of stars do not vary linearly with their 
masses but rather as $M^{3.5}$ or so. Thus a region of high luminosity could 
equally well be due to a large number of low mass stars or to a much fewer 
number of high mass stars, with it in general being an extremely complicated 
matter to try to extract out a galactic matter distribution from a galactic 
luminosity distribution, and perhaps especially so in the star forming spiral 
arm regions prevalent in our galaxy sample. Thus what theory gives us is the 
integration of the true linear potential $V^{t}(r)=\gamma^{t}c^2r/2$ over the 
true matter distribution $\Sigma(R)$. (By 'true' we mean that all nucleons have 
the same $\gamma^{t}$ - though it is in principle possible for protons and 
neutrons to even have different fundamental $\gamma$ parameters - and that for 
weak gravity the total linear term is an extensive function of the number of 
nucleons as measured by $\Sigma(R)$.) Without needing to specify any particular 
form for $\Sigma(R)$, the general analysis of Mannheim (1995a) then yields for 
an infinitesimally thin disk with this $\Sigma(R)$ the net galactic potential
\begin{eqnarray}
rV^{\prime}(r)=
\pi\gamma^{t} c^2r \int_0^{\infty}dR R \Sigma(R) (r^2-R^2) \int_0^{\infty}
dk kJ_1(kr)J_0(kR)                                                   
\label{Eq. (3.1)}
\end{eqnarray}
whose asymptotic $r \rightarrow \infty$ limit is
\begin{eqnarray}
rV^{\prime}(r) \rightarrow
2\pi\gamma^{t} c^2r \int_0^{\infty}dR R \Sigma(R)=\gamma^{t} c^2rN^{t}                    
\label{Eq. (3.2)}
\end{eqnarray}
where $N^{t}$ is the true number of nucleons in the galaxy, a number which in 
principle could differ from the number $N^{*}$ (as multiplied by the total number of
nucleons in the sun) found in the fits. Now $\Sigma(R)$ is not directly measured, 
rather only the luminosity distribution $I(R)$ is detectable, with the relation 
between these two distributions being given as $\Sigma(R)=\mu(R)I(R)$ where 
$\mu(R)$ should be identified as the local mass to light ratio. In terms of this 
local $\mu(R)$ we may rewrite the true total linear potential coefficient of 
the galaxy as
\begin{eqnarray}
\gamma^{t}_{gal}=\gamma^{t}N^{t}
=2\pi\gamma^{t} \int_0^{\infty}dR R \mu(R)I(R)
\label{Eq. (3.3)}
\end{eqnarray}
\noindent
Comparing Eqs. (\ref{Eq. (3.1)}) and (\ref{Eq. (3.3)}) with the ones we 
actually used in the fitting, viz.
\begin{eqnarray}
rV^{\prime}(r)=
{\pi\gamma^{*} rGM \over \beta^{*}L}
\int_0^{\infty}dR R I(R) (r^2-R^2) \int_0^{\infty}dk kJ_1(kr)J_0(kR)                                                   
\label{Eq. (3.4)}
\end{eqnarray}
and
\begin{eqnarray}
\gamma_{gal}=\gamma^{*}N^{*}
={ 2\pi\gamma^{*}GM \over c^2\beta^{*}L} \int_0^{\infty}dR R I(R)
\label{Eq. (3.5)}
\end{eqnarray}
indicates that the parameter $\gamma^{*}$ is really serving as a $\mu(R)$ 
dependent average of $\gamma^{t}$. Since the dependence of $\mu(R)$ on position 
and on given galaxy is not currently known, the variation of effective 
$\gamma^{*}$ with galaxy (or even with the nuclear fusion dependent neutron to 
proton abundance ratio if $\gamma^{n}\neq\gamma^{p}$) is thus also not known. 
However, while the hidden dependence of the fits on $\mu(R)$ represents a 
currently undeterminable effect which prevents us from fully assessing the 
significance of the trend we find for the effective $\gamma^{*}$, nonetheless, 
we regard it as extremely unlikely, remote even, that such dependence on 
$\mu(R)$ could actually account for the enormous variation of effective 
$\gamma^{*}$ with luminosity that is found, or that $\mu(R)$ could possibly 
vary in just the right way in each and every galaxy to make the total effective 
galactic linear potential always come out universal. Thus as such we must 
conclude that the standard Newtonian analysis leads us to an expectation (viz. 
at least somewhat close to universal effective $\gamma^{*}$) which is not at all 
supported by the fits, and in this respect the conformal theory would appear to 
fail to provide an acceptable explanation of rotation curve systematics.
Moreover, given the imposition of universal $\gamma^{*}$, the only way that the 
model of Eqs. (\ref{Eq. (2.2)}) and (\ref{Eq. (2.3)}) could then avoid actually 
even being excluded by the available rotation curve data altogether would be if 
the value of such a universal $\gamma^{*}$ were altogether smaller than any of 
the fitted values for $\gamma^{*}$ found in our fits; and then of course the 
impact of the linear potential term on the rotation curve data would in and of 
itself be way too small to account for any deviation of the data from the 
standard luminous Newtonian expectation at all. (This would incidentally not 
actually make conformal gravity wrong, just somewhat difficult to test - since 
for small enough $\gamma^{*}$ the theory would still enjoy the same Newtonian 
structure as that present in standard gravity.) 

Despite the fact that the model based on the use of Eqs. (\ref{Eq. (2.2)}) and 
(\ref{Eq. (2.3)}) does thus fail to fully account for rotation curve systematics, 
nonetheless, it has uncovered a pattern which the data do respect, namely 
fitting with a universal total galactic linear potential. Thus the fits of 
Figures (1) and (2) reveal that the data do admit of a linear potential, only 
one which, curiously and intriguingly, is normalized with a strength which is 
independent of the amount of matter in each galaxy. Given this lack of 
dependence on particular galaxy, and given the fact that the numerical value of 
the fitted $\gamma_{gal}$ is very close to the inverse Hubble radius, it 
was thus suggested (Mannheim 1995b) that the linear potential needed for the 
fits come not from within each galaxy at all, but rather that it come from the 
effect of the rest of the galaxies in the universe on each given galaxy, an 
effect which is immediately universal and immediately parameterized by a 
cosmological scale. (Indeed, in a theory with linear potentials, i.e. with 
potentials which grow rather than fall with distance, the familiar, purely 
local, Newtonian assumption that we treat galaxies as isolated systems 
is no longer reliable, with each galaxy being strongly influenced by the linear 
potentials of all of the other galaxies in the universe.) Moreover, it was even
shown (Mannheim 1995b) that the local effect of the (explicitly general 
relativistic) global Hubble flow on 
individual galaxies was actually precisely of such universal linear form with
fits then being found which are of a quality comparable with the ones presented 
here, and we refer the reader to Mannheim's paper for further details. While
this new global cosmological view of rotation curves of course needs to be 
explored further, we note that as far as the present paper is concerned, we see 
that our work here has, for its part, uncovered a systematic and completely 
unanticipated pattern in the rotation curve data, a pattern which, while
completely at variance with standard non-relativistic Newtonian reasoning,
would nonetheless still appear to have the capacity to be very instructive.

The authors would like to thank K. Begeman, A. Broeils and C. Carlson for 
helpful comments. J. Kmetko would like to thank W. Stwalley for the kind 
hospitality of the Department of Physics at the University of Connecticut which 
hosted him at its Research Experience for Undergraduates Program in summer 1995.
This work has been supported in part by the Department of Energy under grant No. 
DE-FG02-92ER40716.00.
\vfill\eject

\vfill\eject

{\bf Table (1) Input Data}
\medskip

$$
\begin{array}{cccccccc}

 Galaxy &  Distance & Luminosity & R_0 & M_{HI} &
Photometry & Velocities &
HI~Density \\

 {}& (Mpc) & (10^9L_{B \odot}) & (kpc) & (10^9M_{\odot}) &
&
&
\\

{}&{}&{}&{}&{}&{}&{}&{} \\ 

 DDO~\phantom{1}
        154 &        \phantom{0}4.00&   \phantom{0}0.05&   0.50&
\phantom{0}0.27&
1 &1&
       1 \\

 DDO~\phantom{1}
        170 &        12.01&   \phantom{0}0.16&   1.28&
\phantom{0}0.45&
2 &2&
       2   \\

 NGC~1560&        \phantom{0}3.00&   \phantom{0}0.35 &
1.30&\phantom{0}0.82 &3&3&3   \\

 NGC~3109&        \phantom{0}1.70&   \phantom{0}0.81 &   1.55&
\phantom{0}0.49&
4 &4&
       4\\

 UGC~2259&        \phantom{0}9.80&   \phantom{0}1.02 &   1.33&
\phantom{0}0.43&
5 &5&
       5 \\

 NGC~6503&        \phantom{0}5.94&    \phantom{0}4.80 &  1.73&
\phantom{0}1.57&
6 &7&
       6 \\

 NGC~2403&        \phantom{0}3.25&  \phantom{0}7.90 &    2.05&
\phantom{0}3.10&
6,8 &7&
       6  \\

 NGC~3198&        \phantom{0}9.36&  \phantom{0}9.00&2.72
&\phantom{0}5.00& 6,8 &7&
       6 \\

 NGC~2903&       \phantom{0}6.40&    15.30 &   2.02&
\phantom{0}2.40&
6,8 &7&
       6  \\

 NGC~7331&        14.90&   54.00 &   4.48& 11.30&
8&
7 &7  \\

 NGC~2841&        18.00&    73.60 &  4.55&   15.80&
8 &7& 
7
 
\end{array}
$$

\vfill\eject

{\bf Table (2) Output Parameters for $\gamma_{gas}=\gamma^{*}$}
\medskip
$$
\begin{array}{cccccc}

Galaxy& 
(M/L)_D&\gamma^{*}& \gamma_{gal}& 
(N^{*}\gamma^{*}/L)_D&
\gamma^{*} R^2_0/\beta^{*}   \\

 & (M_{\odot}L_{B\odot}^{-1})&
(10^{-40}cm^{-1})& (10^{-30}cm^{-1})
& (10^{-39}cm^{-1}L_{B \odot}^{-1}) & \\

{}&{}&{}&{}&{}&{} \\

DDO~\phantom{1} 
        154 &      
2.78&
52.80 &       2.82&  
14.68 &      0.085 \\

DDO~\phantom{1} 
        170 &      
7.29&
15.72 &       2.96 &  
11.46 &       0.166 \\

NGC~1560&       
3.48&
18.74 &       4.97 &  
\phantom{0}6.53 &        0.206 \\

NGC~3109&      
0.37&
30.74 &       5.14&  
\phantom{0}1.14 &      0.476 \\

UGC~2259&       
4.36&
\phantom{0}8.53 &       4.42 &  
\phantom{0}3.72 &     0.097 \\

NGC~6503&          
3.14&
\phantom{0}2.24 &       4.02 &  
\phantom{0}0.70&        0.043 \\

NGC~2403&      
2.06&
\phantom{0}2.66 &       5.58 &  
\phantom{0}0.55&        0.072 \\

NGC~3198&       
4.19&
\phantom{0}0.92 &       4.15 &  
\phantom{0}0.39&      0.044 \\
                                                            
NGC~2903&       
3.65&
\phantom{0}1.28 &       7.60 &  
\phantom{0}0.47&  0.033 \\
                                                            
NGC~7331&     
5.31&
\phantom{0}0.42 &       7.13 &  
\phantom{0}0.22&       0.055 \\

NGC~2841&    5.79& 
\phantom{0}0.19 &       6.65 &  
\phantom{0}0.11&   0.026                                                             
                                     
\end{array}
$$

\vfill\eject

{\bf Table (3) Output Parameters for $\gamma_{gas}=0$}
\medskip
$$
\begin{array}{cccccc}

Galaxy& 
(M/L)_D&\gamma^{*}& \gamma_{gal}& 
(N^{*}\gamma^{*}/L)_D&
\gamma^{*} R^2_0/\beta^{*}   \\

 & (M_{\odot}L_{B\odot}^{-1})&
(10^{-40}cm^{-1})& (10^{-30}cm^{-1})
& (10^{-39}cm^{-1}L_{B \odot}^{-1}) & \\

{}&{}&{}&{}&{}&{} \\

DDO~\phantom{1} 
        154 &      
1.85&
261.02 &       2.41&  
48.30 &      0.421 \\

DDO~\phantom{1} 
        170 &      
6.50&
\phantom{0}27.15 &       2.82 &  
17.65 &       0.287 \\

NGC~1560&       
3.02&
\phantom{0}42.66 &       4.50 &  
           12.87 &        0.465 \\

NGC~3109&      
0.33&
139.08 &       3.69&  
\phantom{0}4.55 &      2.157 \\

UGC~2259&       
4.32&
\phantom{00}9.80 &       4.31 &  
\phantom{0}4.23 &     0.112 \\

NGC~6503&          
3.09&
\phantom{00}2.63 &       3.90 &  
\phantom{0}0.81&        0.050 \\

NGC~2403&      
2.01&
\phantom{00}3.43 &       5.45 &  
\phantom{0}0.69&        0.093 \\

NGC~3198&       
4.14&
\phantom{00}1.09 &       4.08 &  
\phantom{0}0.45&      0.052 \\
                                                            
NGC~2903&       
3.64&
\phantom{00}1.36 &       7.54 &  
\phantom{0}0.49&  0.035 \\
                                                            
NGC~7331&     
4.33&
\phantom{00}0.54 &       7.68 &  
\phantom{0}0.23&       0.071 \\

NGC~2841&    6.01& 
\phantom{00}0.20 &       6.54 &  
\phantom{0}0.12&   0.027                                                             
                                     
\end{array}
$$

\vfill\eject   

{\bf Figure Captions}
\medskip

Figure (1). The calculated rotational velocity curves associated with the
linear potential theory for each of the 11 galaxies in the sample fitted 
by varying $\gamma^{*}$ while holding  $\gamma_{gas}/\gamma^{*}$ fixed. In 
each graph the bars show the data points with 
their quoted errors, the full curve shows the overall theoretical velocity 
prediction (in km sec$^{-1}$) as a function of distance from the center of each 
galaxy (plotted in units of $R/R_0$ where each time $R_0$ is each galaxy's own 
scale length), while the dashed and dash-dotted curves show the velocities that 
the Newtonian and linear potentials would then separately produce.

\medskip

Figure (2). The calculated rotational velocity curves associated with the
linear potential theory for each of the 11 galaxies in the sample fitted 
by varying $\gamma^{*}$ while holding  $\gamma_{gas}$ fixed. In 
each graph the bars show the data points with 
their quoted errors, the full curve shows the overall theoretical velocity 
prediction (in km sec$^{-1}$) as a function of distance from the center of each 
galaxy (plotted in units of $R/R_0$ where each time $R_0$ is each galaxy's own 
scale length), while the dashed and dash-dotted curves show the velocities that 
the Newtonian and linear potentials would then separately produce.

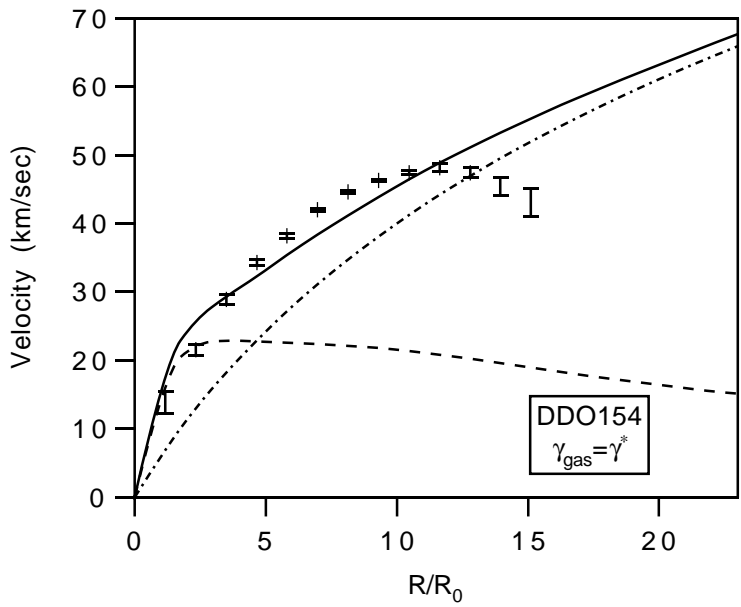
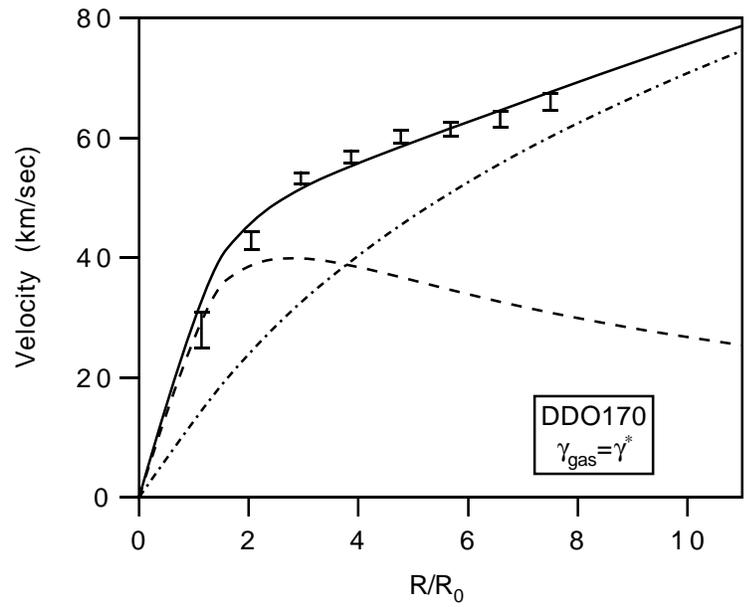
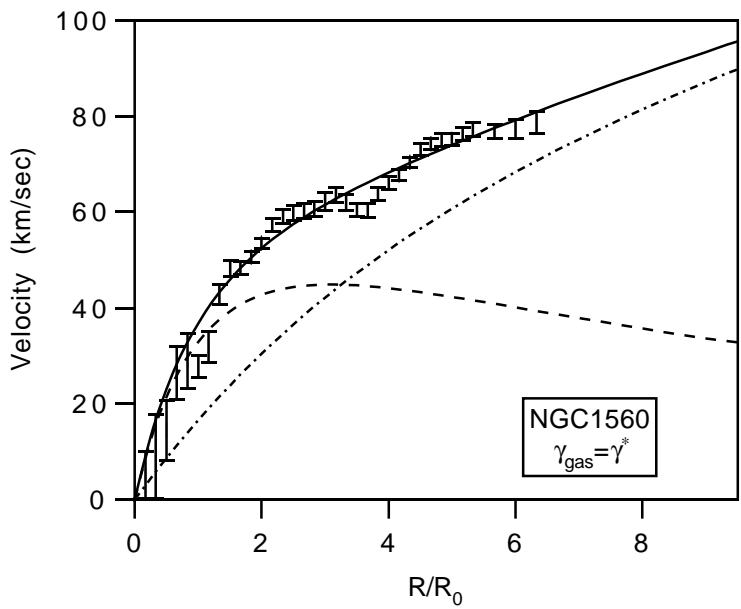
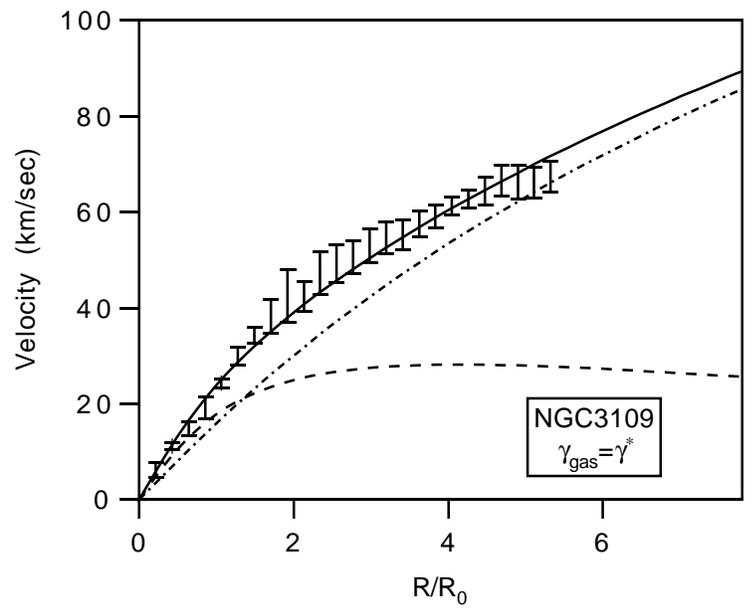
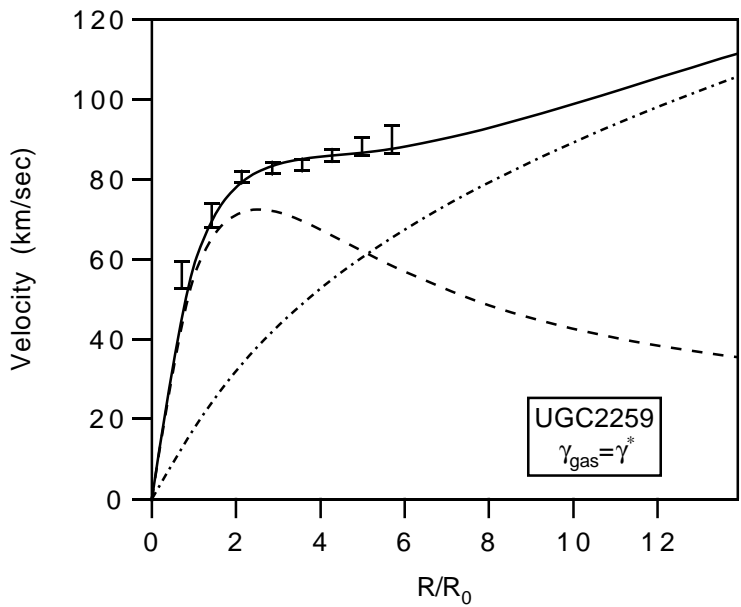
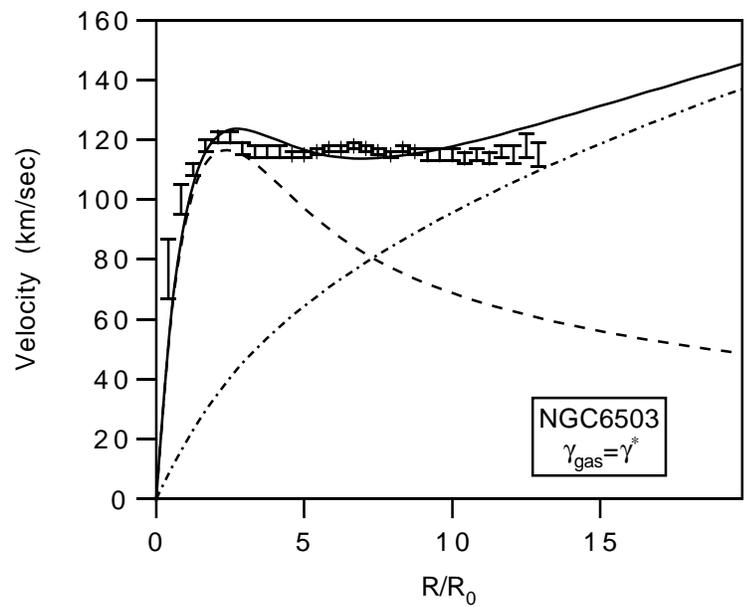
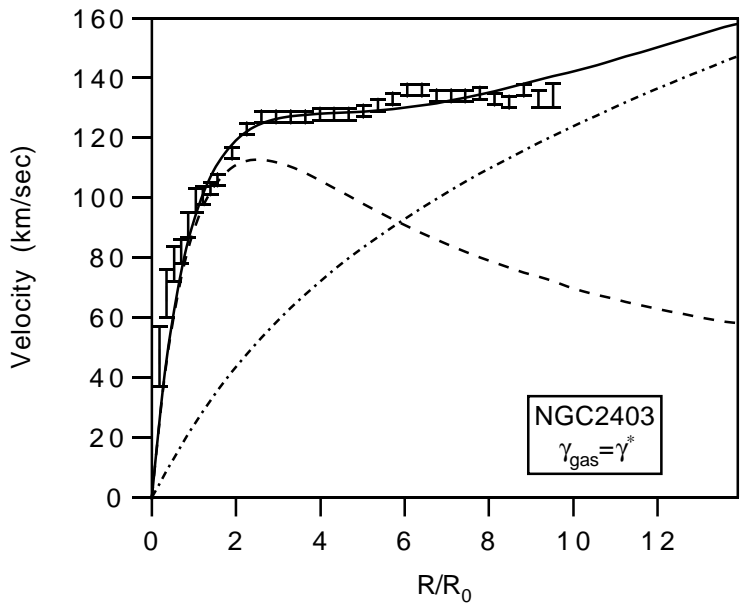
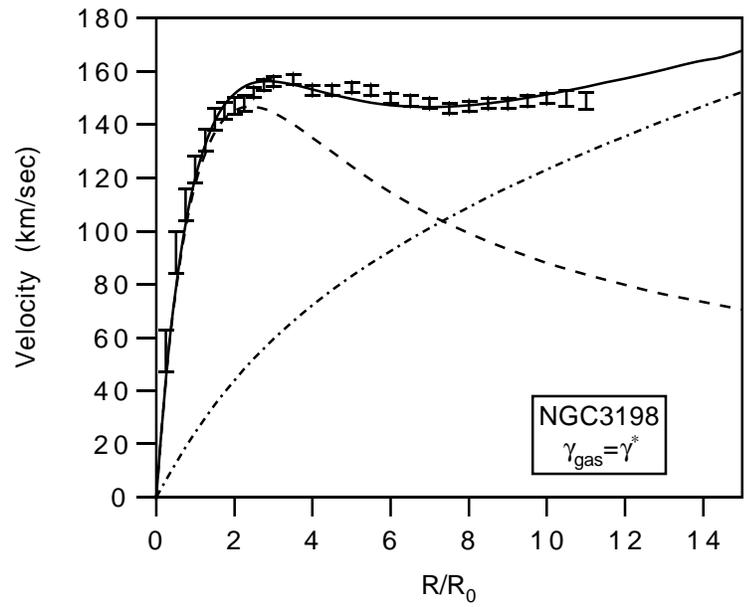
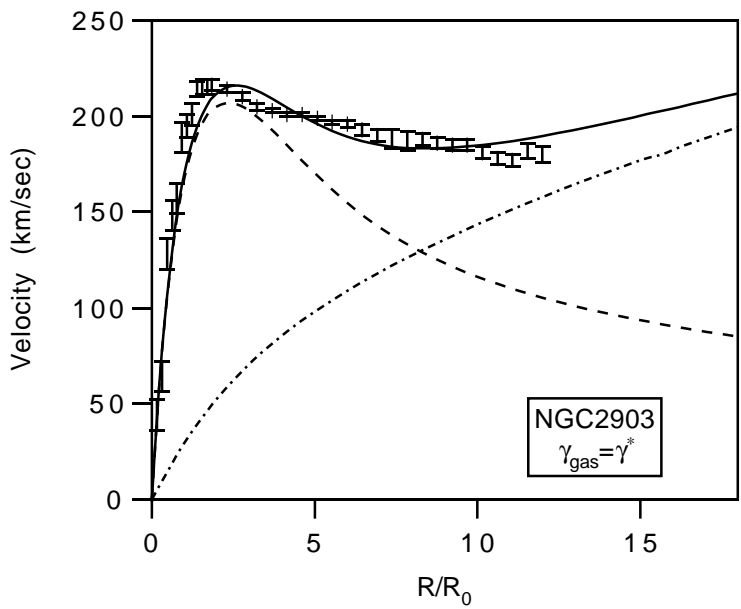
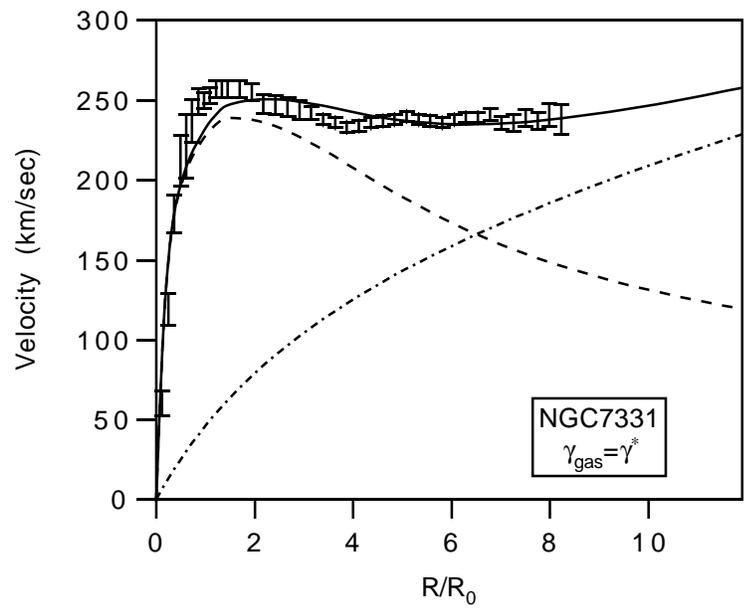
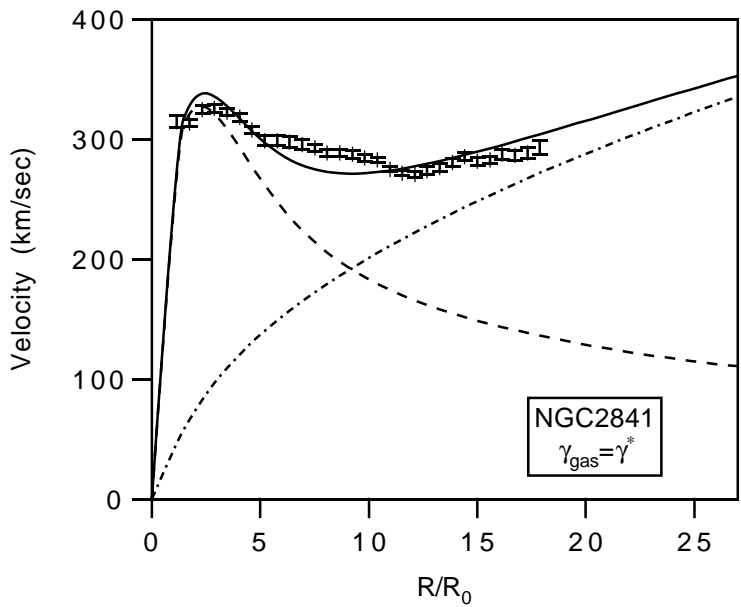
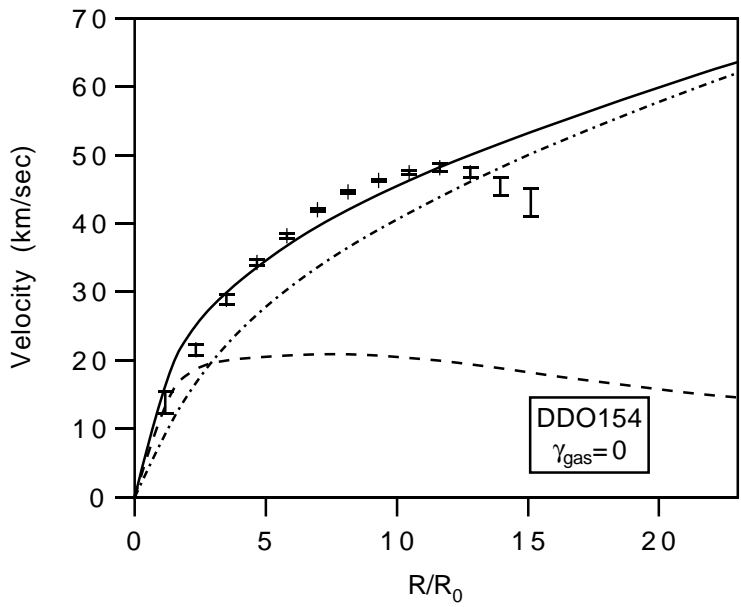
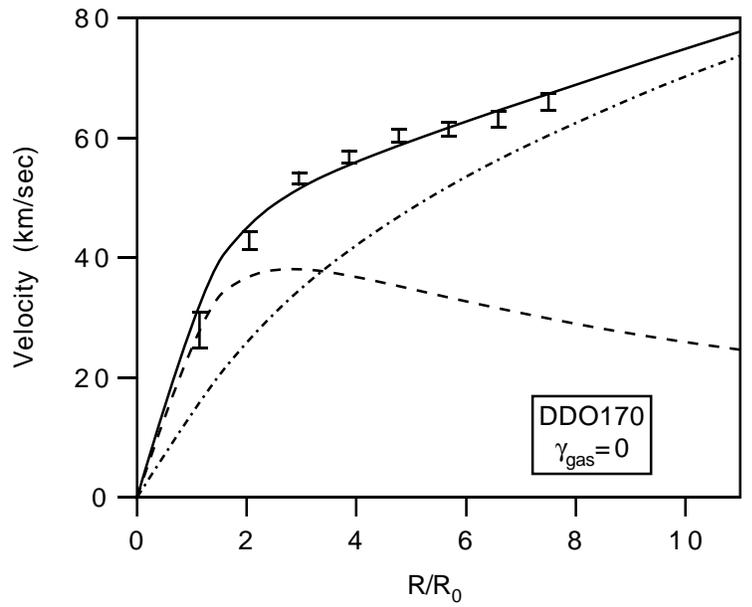
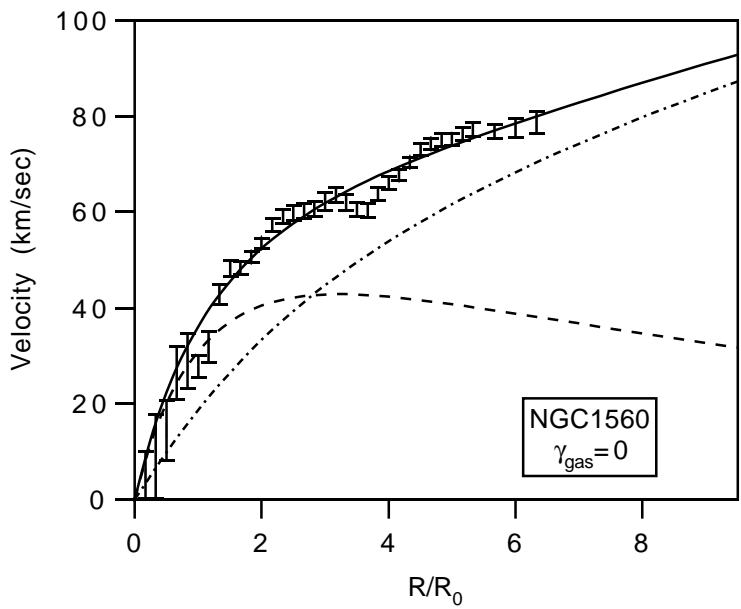
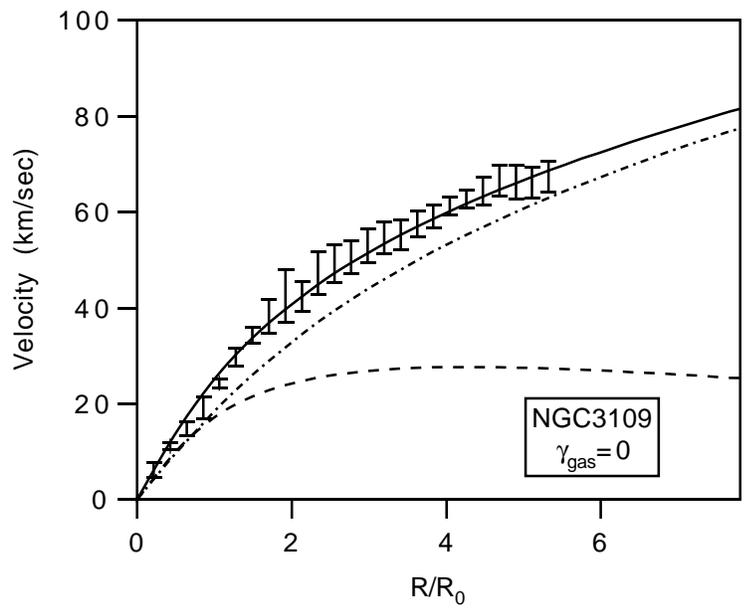
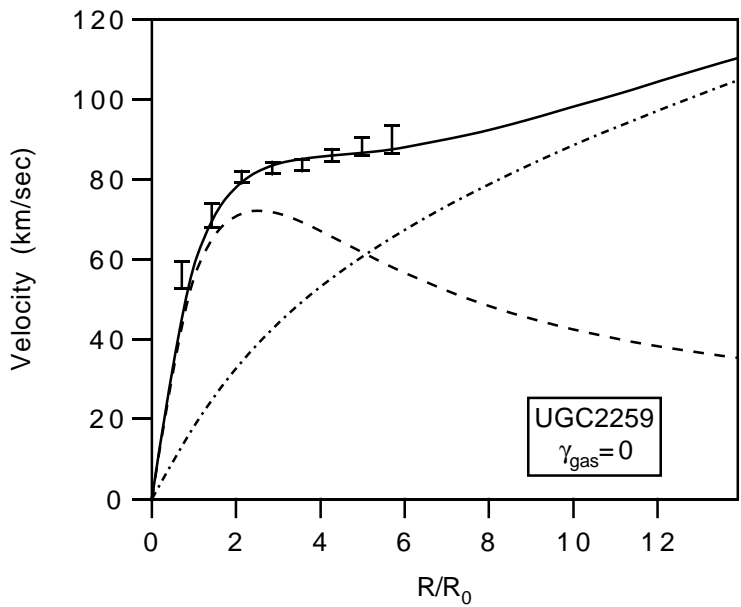
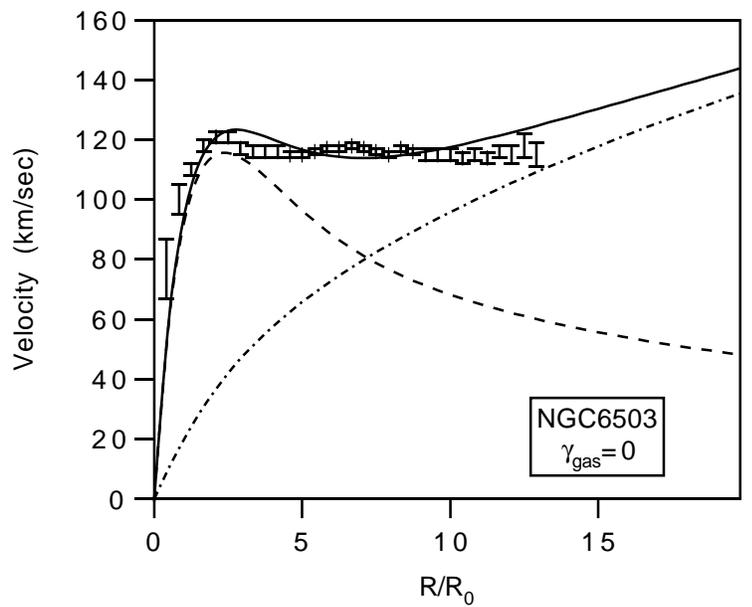
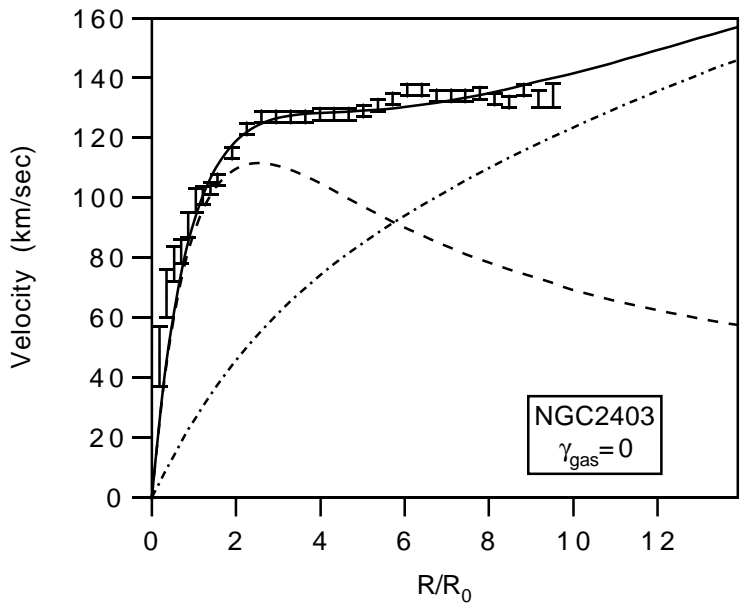
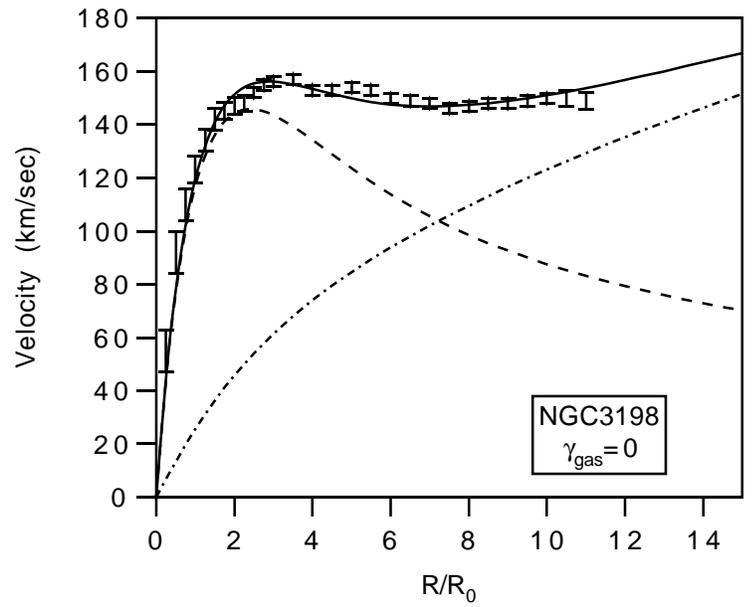
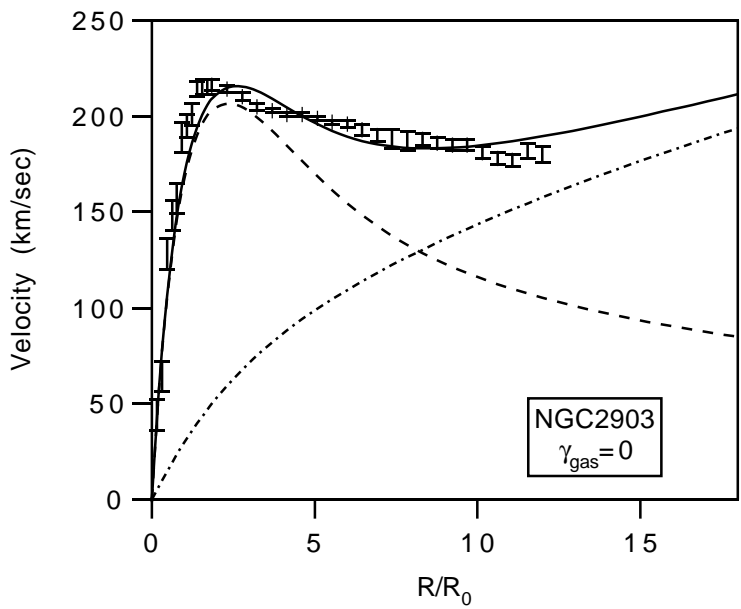
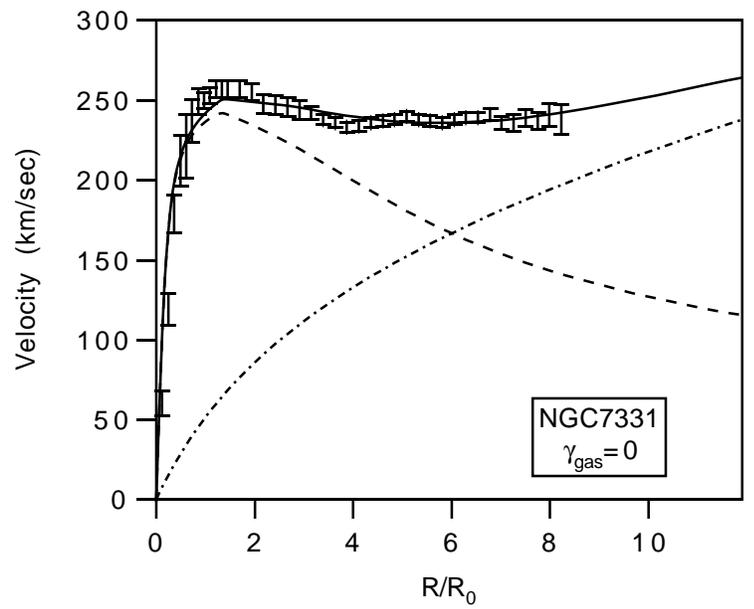
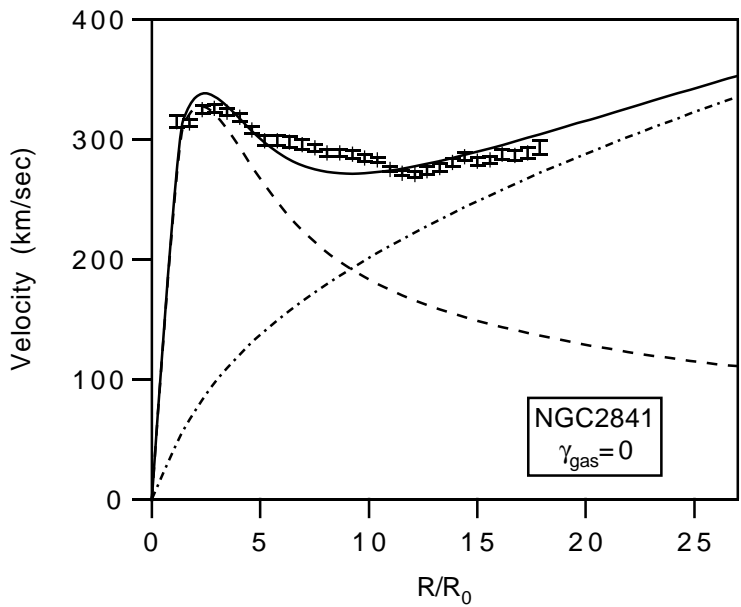
\end{document}